\documentstyle[epsfig,a4]{article}

\begin{document}

\renewcommand{\thesection}{\arabic{section}.}
\renewcommand{\figurename}{\small Fig.}
\renewcommand{\theequation}{\arabic{section}.\arabic{equation}}

\newcommand{\eqreset}{\setcounter{equation}{0}}

\newcommand{\gtrsim}{
\,\raisebox{0.35ex}{$>$}
\hspace{-1.7ex}\raisebox{-0.65ex}{$\sim$}\,
}

\newcommand{\lesssim}{
\,\raisebox{0.35ex}{$<$}
\hspace{-1.7ex}\raisebox{-0.65ex}{$\sim$}\,
}

\begin{flushleft}

{\small J. Phys. A: Math. Gen. {\bf 29} (1996) 2349--2364. Printed in Hamburg}
\vspace{2cm}

{
\Large\bf
Bloch-wall phase transition in the spherical model
}      

\vspace{0.1cm}

\end{flushleft}

\begin{flushright}
\parbox[t]{11cm}{

D A Garanin
\renewcommand{\thefootnote}{\fnsymbol{footnote}}
\footnotemark[2]
\vspace{0.1cm}

\small

I. Institut f\"ur Theoretische Physik, Universit\"at Hamburg,
Jungiusstr. 9, \\ D-20355 Hamburg, Germany\\ 
\vspace{0.2cm}

Received 24 August 1995, in final form 15 December 1995

\vspace{0.4cm}

{\bf Abstract.} \hspace{1mm}
The temperature-induced second-order phase transition from Bloch to linear 
(Ising-like) 
domain walls in uniaxial ferromagnets is investigated for the model of 
$D$-component classical spin vectors in the limit $D\to\infty$.
This exactly soluble model is 
equivalent to the standard spherical model in the homogeneous case, but
deviates from it and is free from unphysical behavior in a
general inhomogeneous situation. 
It is shown that 
the thermal fluctuations of the transverse magnetization in the wall 
(the Bloch-wall order parameter) result in the diminishing of the 
wall transition temperature $T_B$ in comparison to its 
mean-field value, thus favouring the existence of linear walls. 
For  
finite values of $T_B$ an additional anisotropy in the basis plane 
$x,y$ is required; in purely uniaxial ferromagnets a domain wall 
behaves like a 2-dimensional system with a continuous spin symmetry and 
does not order into the Bloch one. 
} 
\vspace{0.8cm}

\end{flushright}

\renewcommand{\thefootnote}{\fnsymbol{footnote}}
\renewcommand{\footnoterule}{\rule{0cm}{0cm}}
\footnotetext[2]{E-mail: garanin@physnet.uni-hamburg.de}

\section{Introduction}

The spherical model proposed in 1952 by Berlin and Kac \cite{berkac52} 
(see also \cite{joy72pt,khokhopassch92})
is extensively used up to now as the only exactly soluble model 
describing the phase transition in 3-dimensional magnetic systems. 
In 
contrast to the mean field approximation (MFA), the spherical model 
describes, in a simplified manner, the thermal fluctuations of spins, which 
can be taken into account exactly due to their Gaussian nature. 
The 
technique for the consideration of {\em inhomogeneous} systems described by 
the spherical model was developed by Barber and Fisher 
\cite{barfis73} for the surface-induced inhomogeneity in 
layered magnetics and elaborated by Abraham and Robert 
\cite{abrrob80} for the problem of phase separation 
(i.e., the domain wall (DW) formation) in the spherical model. 
Later the 
inhomogeneous states of the bounded spherical model induced by 
antiperiodic \cite{sinpatfis86} and twisted \cite{all95} boundary 
conditions were investigated.

The results obtained for inhomogeneous states of the spherical model 
posess some unexpected features. The phase transition temperature 
$T_c$ of a 4-dimensional ferromagnetic slab consisting of $N\gg 1$ layers 
turns out to be higher than that of the bulk
in the case of {\em free-edge} boundary conditions  
\cite{barfis73}. 
The two-domain state induced by the magnetic field 
$\pm H$ in two half-spaces is characterized by the DW width diverging 
and the domain magnetization vanishing in the limit $H\to 0$, 
i.e., in contrast to the underlying Ising model the spherical model does not 
exhibit the phase separation \cite{abrrob80}.
As was argued already in \cite{barfis73}, 
such features are the consequence of the global spin constraint
\cite{berkac52}, which obviously becomes not so good 
in the inhomogeneous case.

Another version of the spherical model --- 
the model of isotropic $D$-component 
classical spin vectors in the limit $D\to \infty$ \hspace{1mm} ---
was proposed by 
Stanley \cite{sta68,sta74}, who showed that in the homogeneous case 
this model is equivalent to the spherical model by Berlin and 
Kac \cite{berkac52}. 
The normalization condition for a spin on a lattice 
site $i$, $|{\bf m}_i|=1$, becomes, in the case $D\to \infty$, 
very similar to 
the global spin constraint, which is the reason of the equivalence of 
the two models. 
However, since the spin normalization condition does not mix 
the spin variables on different lattice sites, the $D$-vector model by 
Stanley with $D\to\infty$ is more physically appealing than the original 
spherical model. 
Moreover, the two models become non-equivalent in a 
general inhomogeneous situation, where, as was shown by Knops 
\cite{kno73}, the $D=\infty$ model corresponds to some generalization of 
the spherical model using a local spin constraint. 
This idea was 
substantiated in the work by Costache {\em et al} 
\cite{cosmazmih76}, who calculated the Curie temperature $T_c(N)$ of a 
ferromagnetic film of $N\gg 1$ layers with free boundary conditions in 4 
dimensions using a set of independent spin constraints in each layer. 
The numerically calculated values of $T_c(N)$ monotonically increase 
with $N$ to the 
bulk value $T_c(\infty)$, which is physically expected and differs from 
the result of Barber and Fisher \cite{barfis73} for the standard 
spherical model.

The further advantages of the $D$-vector 
model are the possibilities of consideration of finite-$D$ and 
anisotropic systems. 
The latter is important, in particular, for the 
calculation of finite-size corrections to $T_c$ of ferromagnetic films 
mentioned above 
in the actual 3-dimensional case. 
Since such a film with $N<\infty$ is a 
2-dimensional system, the $T_c$-corrections are finite only in the 
presence of the stabilizing uniaxial anisotropy \cite{gar96jpal}.
In spite of its advantages in comparison to the 
standard spherical model, the $D$-vector model with $D\gg 1$ was much 
less used than mentioned. 
As exclusions one can cite the works by 
Abe and Hikami \cite{abe7273,abehik7377} and by 
Okabe and Masutani \cite{okamas78} dealing with the $1/D$ expansion for 
3-dimensional systems and the work by Okamoto \cite{oka70} where the 
uniaxial spherical model with a transverse field was considered.

It should be noted that practically all the researchers dealing with the 
{\em isotropic} $D$-vector model used the designations $N$ or $n$ 
instead of the original $D$. 
Such a modification is, however, 
not justified in a general anisotropic case, where the number $n$ 
of the relevant order parameter components determining the 
symmetry and thus the universality class of a system can be 
different from $D$. 
As an example one can consider the rather 
general ``$n$-$D$ model'' \cite{garlut84d,gar96prb} 
having the first $n\leq D$ components 
coupled by the exchange interaction with the equal 
strength and the rest $D-n$ components ``free''. 
Among 
realizations of the $n$-$D$ model are, in particular, the $x$-$y$ 
model ($D=3$, $n=2$) and the plane rotator one ($n=D=2$) belonging  
to the same universality class determined by $n$ but having 
different values of $T_c$ depending on both $n$ and $D$. 
Correspondingly, in a general case the $1/n$ expansion of the critical 
indices is not the same as the $1/D$ expansion of non-universal 
quantities.

The general qualitative result of  \cite{abrrob80}, the absence of 
the phase 
separation in the spherical model (but not the disappearance of the 
domain magnetization !), can be explaned by the fact that this 
model behaves in the bulk like the {\em isotropic} $D=\infty$ model
\cite{sta68,garlut84d}, which 
obviously exhibits no phase separation. 
For the $D$-vector model the 
separation of a specimen into domains with opposite magnetizations by 
domain walls of a finite width requires an easy-axis anisotropy, which 
makes the intermediate orientation of the magnetization in the wall 
energetically unfavourable in comparison to that in domains. 
Clearly, this actual situation cannot be treated either with the help of the 
spherical model in its standard formulation \cite{berkac52}, or
with the improved version \cite{cosmazmih76,angbuncos81}, which is 
equivalent to the {\em isotropic} $D=\infty$ model 
in the general inhomogeneous case.

The problem
arising here --- the study of the influence of thermal 
fluctuations on 
the domain wall structure --- is not only important for comparing the 
properties of different versions of the spherical model. 
The physics of domain walls at 
elevated temperatures is itself rather interesting and unexplored area, 
whereas since the seminal work by Landau and Lifshitz \cite{lanlif35} 
the majority of researchers have addressed the zero-temperature 
statics and dynamics of DWs based on the assumption of the constant 
magnitude of the magnetization in the wall.

The first theoretical investigation of the temperature variation of the 
structure of domain walls is due to 
Bulaevskii and Ginzburg \cite{bulgin64}, 
who with the help of the phenomenological version of MFA,
using a macroscopic Landau free energy in the vicinity of $T_c$, 
predicted a phase transition from Bloch to linear DWs 
in uniaxial ferromagnets at some $T_B<T_c$. 
Qualitatively this phase transition can be explaned by 
the fact that the spins in the center of a Bloch wall, which are 
forced to lie perpendicular to the easy axis, experience a 
molecular field smaller than in domains and hence order at some 
temperature 
$T_B$ less than $T_c$, which leads to linear (Ising-like) walls (LW)
in the region $T_B<T<T_c$. 
For 
ferromagnets whose anisotropy energy is much smaller than the exchange 
interaction, the LW temperature region is narrow.

Complementary, the transition from Bloch to linear walls 
at $T=0$ depending on the anisotropy was investigated 
by van den Broek and Zijlstra \cite{brozij71} 
with numerical methods. 
It was found that LWs are realized if the ratio 
of the anisotropy energy to the exchange one exceeds 2/3; the DW width 
$\delta$, is in this case, comparable with the lattice spacing $a_0$. 
Later this transition was discovered by Sarker, 
Trullinger and Bishop \cite{sartrubis76} in the framework of a formal 
soliton theory independently of Bulaevskii and Ginzburg. 
The problem was addressed also by 
Niez and Lajzerowicz \cite{nie76,lajnie79}, 
where the factor 2/3 mentioned above was calculated analytically.

The first indirect experimental evidence for the transition from Bloch to 
linear domain walls was obtained from the optical observations of the 
temperature dependence of the period of the domain structure in YFeO$_3$ 
just below $T_c$ \cite{szymazpio79}. 
Lately the LWs
were observed in the dynamical susceptibility 
experiments on the low-temperature 
ferromagnets ${\rm GdCl_3}$ \cite{grakoe89} and ${\rm LiTbF_4}$ 
\cite{koegrasesfer90}. 
In  \cite{pangarrus90,gar91llbedw} the DW 
mobility was calculated in the whole temperature range, which exibited 
a deep minimum at $T_B$. 
Such a minimum was 
observed recently in the dynamical susceptibility experiments 
on the high-temperature Ba and Sr hexaferrites 
\cite{koegarharjah93,harkoegar95}.

Recent experiments also provided evidence of strong 
fluctuational effects about the DW phase transition. 
The transition 
temperature $T_B$ was substantially lower than its mean-field estimate, 
and 
the critical index $\beta_B$ of the Bloch-wall order parameter (the 
transverse magnetization in the centre of the wall) was about 0.1 in 
contrast to the MFA value 1/2. 
Such strong fluctuations are actually not 
surprising since a domain wall is a 2-dimensional object. 
The 
analysis by Lawrie and Lowe \cite{lawlow81} making use of 
renormalization-group arguments
has led to the clear result that a domain wall in a {\em biaxial} 
ferromagnetic model having 
an additional anisotropy in the $x,y$ plane (which is usually the 
effective one due to the magnetostatic field \cite{lanlif35}) belongs to 
the universality class of the 2-dimensional Ising model, and hence one 
can expect $\beta_B=1/8$. 
In contrast, in a purely uniaxial 
model without the dipole-dipole interaction a 
domain wall behaves like a 2-dimensional plane rotator model and can 
show only the Kosterlitz-Thouless phase transition without ordering to a 
Bloch wall.

The absense of the long-range order 
(i.e., the transverse magnetization component) 
in a domain wall in a purely uniaxial ferromagnet 
can be demonstrated with the help of the 
linear spin-wave theory.
The thermal disordering of Bloch walls is due to the 
so-called Winter magnons \cite{win61}, the excitations 
localized on the domain wall with the dispersion law 
$\varepsilon_q^2 \propto A q^2(A q^2 + K_\perp)$ 
($A$ is the inhomogeneous exchange constant, $K_\perp$ 
is the in-plane anisotropy constant and 
$q$ is a 2-dimensional wavevector). 
The first factor in $\varepsilon_q^2$ 
corresponds to the free translational motion of the wall and the second 
describes the rotation of the magnetization in the center of the wall in 
the $x,y$ plane.
It can be seen that the number of 
Winter magnons, which in the classical case is proportional to 
$\int d^2 q/\varepsilon_q$, diverges logarithmically at 
small $q$, if $K_\perp$ tends to zero. 
Thus, in a purely uniaxial 
ferromagnetic model the linear walls cannot order to the Bloch ones at 
any non-zero temperature. For a small non-zero in-plane anisotropy $K_\perp$
the shift of $T_B$ from its MFA value due to fluctuations 
should be very essential.

The aim of this work is to find an exact solution for the domain wall 
magnetization profile in biaxial ferromagnets at non-zero temperatures 
and their transition temperature $T_B$ in 
the framework of the spherical model in its $D$-vector version. 
Instead 
of modifying the approach of  \cite{barfis73,abrrob80} based on 
the calculation of the partition function with the steepest descent 
method, we will use 
the diagram technique for classical spin systems 
\cite{garlut84d,gar94jsp}. 
This diagram technique,
which is a generalization of the ``Ising part'' of the spin
operator diagram technique of Vaks {\em et al} \cite{vlp67s,izyskr88}, 
makes it possible to 
locate and sum up all the diagrams surviving in the limit $D\to\infty$ 
and can be reformulated for the present purposes for inhomogeneous 
situations.

The approximation obtained by summing up such
diagrams (without going to the limit $D\to\infty$) is the
so-called self-consistent Gaussian approximation (SCGA), which
was first formulated by Horwitz and Callen \cite{horcal61} for
the Ising model ($D=1$). 
SCGA yields rather good results for the
thermodynamic quantities of the Ising
\cite{garlut84i} and the classical Heisenberg ($D=3$)
\cite{garlut86jpf} models on 3-dimensional lattices in the whole 
temperature range and can be of importance for a possible improvement
of the presently obtained results for the domain wall structure in the
spherical limit with regard to systems with finite $D$. 
A detailed 
description of the classical spin diagram technique and SCGA can be 
also found in the recent publication \cite{gar96prb}.

The rest of the paper is organized as follows.
In Section 2 the diagram technique for classical spin systems
and the construction of SCGA in the inhomogeneous case is 
described. 
In Section 3 SCGA is simplified for $D\to\infty$,
and a closed system of equations for magnetization and spin-spin 
correlation function describing the domain wall in the spherical limit is 
derived. 
In Section 4 the magnetization profile of a fluctuating domain 
wall is calculated and the dependence of the 
transition temperature $T_B$ on the anisotropy parameters is analyzed. 
In Section 5 further problems of the DW statics and dynamics at 
elevated temperatures are discussed.

\section{Classical spin diagram technique and SCGA}
\eqreset

The appropriate classical $D$-vector Hamiltonian with biaxially 
anisotropic ferromagnetic exchange interaction can be written in the form
\begin{equation}\label{dham}
{\cal H} = 
- \frac{1}{2}\sum_{ij}J_{ij}
\left(
m_{zi}m_{zj} + \eta m_{yi}m_{yj}
+ \sum_{\alpha=3}^D \eta_\alpha m_{\alpha i} m_{\alpha j}
\right) ,
\end{equation}
where $i,j$ are the lattice sites, 
${\bf m}_i$ is the normalized $D$-component vector, 
$|{\bf m}_i|=1$, the dimensionless ani\-so\-t\-ro\-py factors satisfy 
$\eta_\alpha \leq \eta \leq 1$ and all $\eta_\alpha$ are, 
for simplicity, taken to be equal to each other. 
For $D=3$ equation (\ref{dham}) reduces to the 
anisotropic classical Heisenberg model. 
In the chosen geometry the 
average magnetization in the bulk is directed 
parallel or antiparallel to the 
easy axis $z$, in the centre of a Bloch wall it takes on one of two 
possible orientations along the second easy axis $y$. 
All variables 
describing the magnetization profile of a plane DW are functions of the 
coordinate $x$ only. 
The temperature-normalized molecular field 
$\mbox{\boldmath$\xi$}_i$
acting on a spin on the site $i$ from its neighbours is given by
\begin{equation}\label{mfield}
\mbox{\boldmath$\xi$}_i = 
-\frac{\partial (\beta{\cal H}) }{\partial{\bf m}_i} 
= \beta
\sum_j J_{ij}
\left(
m_{zj}{\bf e}_z + \eta m_{yj}{\bf e}_y
+ \sum_{\alpha=3}^D \eta_\alpha m_{\alpha j} {\bf e}_\alpha
\right) ,
\end{equation}
where $\beta\equiv 1/T$ and  ${\bf e}$ are unit vectors in appropriate 
directions. The mean 
field approximation consists in neglecting fluctuations of the 
molecular field $\mbox{\boldmath$\xi$}_i$; replacing 
$m_{zj}\Rightarrow \langle m_{zj}\rangle$, 
$m_{yj}\Rightarrow \langle m_{yj}\rangle$ and 
$m_{\alpha j}\Rightarrow 0$ in (\ref{mfield}), one arrives at the 
inhomogeneous Curie-Weiss equation 
\begin{equation}\label{cweiss}
\langle {\bf m}_i\rangle = B(\xi_i)\mbox{\boldmath$\xi$}_i/\xi_i ,
\end{equation}
where $B(\xi)$ is the Langevin function. 
For small 
anisotropy ($1-\eta \ll 1$) the magnetization varies slowly on the scale 
of the lattice spacing, and the continuous approximation can be applied to 
(\ref{cweiss}). 
In this case the zero-temperature results of 
Landau and Lifshitz \cite{lanlif35} and the finite-temperature ones of 
Bulaevskii and Ginzburg \cite{bulgin64} for the DW magnetization profile 
are recovered (see below).

\begin{figure}[t]
\unitlength1cm
\begin{picture}(15,8)
\centerline{\epsfig{file=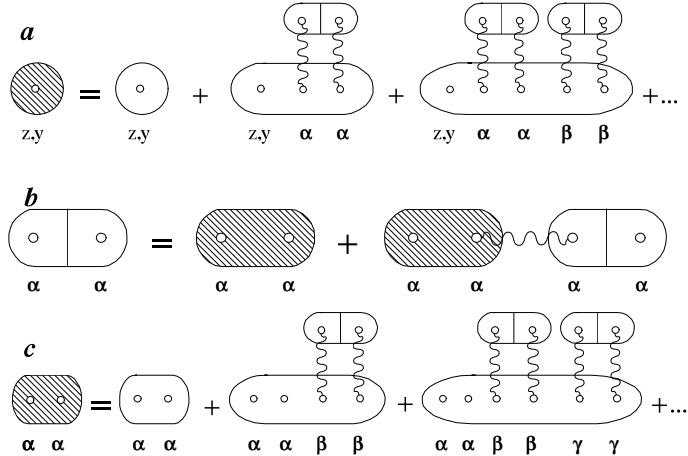,angle=0,width=16cm}}
\end{picture}
%
\caption{ \label{sdwf1}
Self-consistent Gaussian approximation (SCGA) for classical spin 
systems.
(a) and (c): block summations for the renormalized magnetization 
and pair spin cumulant averages; 
(b): Dyson equation for the spin-spin correlation function.
}
\end{figure}

Fluctuations of the molecular field (\ref{mfield}) 
can be taken into account within the framework of a perturbative scheme 
based on the diagram technique for classical spin systems 
\cite{garlut84d,gar94jsp,gar96prb}.
The perturbative expansion of the thermal average of any quantity ${\cal A}$ 
characterizing a classical spin system (e.g., ${\cal A} = {\bf m}_i$) can be 
obtained by rewriting (\ref{dham}) as 
${\cal H} ={\cal H}_0 + {\cal V}_{\rm int}$, 
where ${\cal H}_0$ is the MFA Hamiltonian with the averaged molecular 
field determined by (\ref{cweiss}), and expanding the expression 
\begin{equation}\label{statavr}
\langle {\cal A}\rangle = \frac{1}{{\cal Z}} \int\prod_{j=1}^N d{\bf m}_j
{\cal A} \exp(-\beta {\cal H}), \qquad  |{\bf m}_j|=1 
\end{equation}
in powers of ${\cal V}_{\rm int}$. 
The averages of various spin vector components 
$\alpha,\beta,\gamma,\ldots=1,\ldots,D$
on various lattice sites $i,j,k,\ldots$ with the Hamiltonian 
${\cal H}_0$ can be 
expressed through spin cumulants, $\Lambda_{\ldots}$, (see below) 
in the following way:
\begin{eqnarray}\label{siteavr}
&& 
\langle m_{\alpha i}\rangle_0  =  \Lambda_{\alpha i}, \nonumber   \\
&&
\langle m_{\alpha i}m_{\beta j}\rangle_0 = \Lambda_{\alpha\beta i} \delta_{ij} 
+ \Lambda_{\alpha i} \Lambda_{\beta j},         \\
&& 
 \langle m_{\alpha i}m_{\beta j}m_{\gamma k}\rangle_0  =  
 \Lambda_{\alpha\beta\gamma i} \delta_{ijk} 
+ \Lambda_{\alpha\beta i}\Lambda_{\gamma k} \delta_{ij}  \nonumber \\
&&\qquad\qquad\qquad
+\, \Lambda_{\beta\gamma j}\Lambda_{\alpha i} \delta_{jk}
+ \Lambda_{\gamma\alpha i}\Lambda_{\beta j} \delta_{ki}
+ \Lambda_{\alpha i} \Lambda_{\beta j} \Lambda_{\gamma k} ,   \nonumber
\end{eqnarray}
etc., where $\delta_{ij}$, $\delta_{ijk}$, etc., are the site Kronecker 
symbols equal to 1 for all site indices coinciding with each other and to 
zero in all other cases. 
For one-site averages 
($i=j=k=\ldots$) equation (\ref{siteavr}) reduces to the well-known 
representation of moments through cumulants (semi-invariants), 
generalized for the multi-component case. 
In the graphical language (see, e.g., Fig. \ref{sdwf1}) the decomposition 
(\ref{siteavr}) corresponds to all possible groupings of small circles 
(spin components) into oval blocks (cumulant averages). 
The circles 
coming from ${\cal V}_{\rm int}$ (the ``inner'' circles) are connected pairwise 
by the wavy interaction lines representing the quantity  
$\eta_\alpha\beta J_{ij}$. 
In diagram expressions the 
summation over the coordinates 
$i$ and component indices $\alpha$ of inner circles is carried out. 
One should not take into account disconnected diagrams (i.e., those 
containing disconnected parts with no ``outer'' circles belonging to 
${\cal A}$ in (\ref{statavr})), since these diagrams are compensated for 
by the expansion of the partition function ${\cal Z}$ in the denominator 
of (\ref{statavr}). 
Consideration of numerical factors shows 
that each diagram contains the factor $1/n_s$, where $n_s$ is the number 
of the symmetry group elements of a diagram (see 
(\ref{lalphai}) and (\ref{rencumlsmall}), the symmetry operations do 
not concern outer circles). 
In the homogeneous case 
it is more convenient for practical 
calculations to use the Fourier representation and 
calculate integrals 
over the Brillouin zone rather than lattice sums. 
As the lattice sums 
are subject to the constraint that the coordinates of the circles 
belonging to the same block coincide with each other (due to 
the Kronecker symbols in (\ref{siteavr})), in the Fourier representation 
the sum of wavevectors coming to or going out of any block along 
interaction lines is zero. 
The cumulant spin averages in (\ref{siteavr}) 
can be obtained by differentiating the generating function 
$\Lambda(\xi)=\ln {\cal Z}_0(\xi)$ 
over appropriate components of the dimensionless molecular field 
$\mbox{\boldmath$\xi$}$:
\begin{equation}\label{defcum}
\Lambda_{\alpha_1\alpha_2..\alpha_p}(\mbox{\boldmath$\xi$})=
\frac{\partial ^p\Lambda(\xi)}{\partial \xi_{\alpha_1} 
\partial \xi_{\alpha_2}..\partial \xi_{\alpha_p}}, 
\end{equation}
where 
${\cal Z}_0(\xi)={\rm const}\times \xi^{-(D/2-1)}{\rm I}_{D/2-1}(\xi)$
is the partition function of a $D$-component classical spin and
${\rm I}_\nu(\xi)$ is the modified Bessel function. The two lowest-order 
cumulants which will be needed below can be written  
explicitly as
\begin{eqnarray}\label{cum12}
&&
\Lambda_\alpha(\mbox{\boldmath$\xi$}) 
= B(\xi) \frac{\xi_\alpha}{\xi} ,                         \\
&&
\Lambda_{\alpha\beta}(\mbox{\boldmath$\xi$}) 
= \frac{B(\xi)}{\xi}
\left( \delta_{\alpha\beta} - \frac{\xi_\alpha \xi_\beta}{\xi^2} \right)
+ B'(\xi) \frac{\xi_\alpha \xi_\beta}{\xi^2} , \nonumber
\end{eqnarray}
where $\delta_{\alpha\beta}$ is the spin component Kronecker symbol, 
$B(\xi) = d\Lambda(\xi)/d\xi$ is the Langevin function for $D$-component 
spins and $B'(\xi) \equiv dB(\xi)/d\xi$.
The expressions for the 3- and 4-spin cumulants can be found in 
\cite{gar94jsp}.
It should be stressed that the spin cumulants (\ref{cum12}) appearing in the 
diagrams 
generated originally by the expansion of (\ref{statavr}) in powers of 
${\cal V}_{\rm int}$ (the unrenormalized diargams) simplify, since there is 
only few non-zero components of the molecular field  
$\mbox{\boldmath$\xi$}$ (for a domain wall in the chosen geometry 
$\xi_z$ and $\xi_y$). 
The complete 
form of spin cumulants (\ref{cum12}) is needed, however, for 
the construction of SCGA allowing for the fluctuations of other 
components of the molecular field. 
For Ising systems the classical spin 
diagram technique coincides with the ``Ising part'' of the standard SDT
\cite{vlp67s,izyskr88} and can be used with Brillouin 
functions $B_S$ of a general spin $S$. 
In the book \cite{izyskr88} more technical details
concerning the construction of SDT for Ising systems can be found, 
which play the same role in the present classical SDT. 
\par
The next step beyond MFA is the self-consistent Gaussian approximation 
taking into 
account {\em pair} correlations of the molecular field acting on a given 
spin from its neighbors, which implies the Gaussian statistics of the 
molecular field fluctuations (see Fig. \ref{sdwf1}). 
Since subsequently we are 
going to take the limit $D\to\infty$, only fluctuations of 
the molecular field components with $\alpha=3,\ldots,D$ should be 
taken into account, because their total contribution exceeds that of the 
fluctuations of $z$ and $y$ components by a factor of the order of $D$. 
The diagram sequence represented in Fig. \ref{sdwf1}
is equivalent to a closed system of nonlinear equations 
for the averaged magnetization $\langle {\bf m}_i\rangle$ 
and the correlation function 
$S_{ij}^{\alpha\alpha}\equiv \langle m_{\alpha i} m_{\alpha j}\rangle$
of the spin components with $\alpha=3,\ldots,D$. 
The diagrammatic equation 
in Fig. \ref{sdwf1}a is the generalization of the 
Curie-Weiss equation (\ref{cweiss}) for the magnetization 
(the angle brackets are dropped):
\begin{equation}\label{scgam}
{\bf m}_i = 
\partial \tilde\Lambda(\mbox{\boldmath$\xi$}_i, l_{\alpha i}) 
/\partial \mbox{\boldmath$\xi$}_i 
= \tilde\Lambda_z(\mbox{\boldmath$\xi$}_i, l_{\alpha i}) {\bf e}_z
+ \tilde\Lambda_y(\mbox{\boldmath$\xi$}_i, l_{\alpha i}) {\bf e}_y
\end{equation}
where the (averaged) molecular field $\mbox{\boldmath$\xi$}_i$ is given 
by the expression (\ref{mfield}) without the last term and 
$l_{\alpha i}$ is related to the dispersion of the molecular field 
fluctuations on the site $i$:
\begin{equation}\label{lalphai}
l_{\alpha i} \equiv \frac{1}{2!}\langle \xi_{\alpha i} \xi_{\alpha i}\rangle 
= \frac{1}{2!} \eta_\alpha^2 \beta^2 \sum_{jj'}J_{ij}J_{ij'}
S_{jj'}^{\alpha\alpha}.
\end{equation}
The spin cumulant averages $\tilde\Lambda_{\ldots}$ on a site $i$ 
(see Fig. \ref{sdwf1}a,c)
renormalized by Gaussian fluctuations of the molecular field are given 
by
\begin{equation}\label{rencumlsmall}
\tilde\Lambda_{\ldots} = \Lambda_{\ldots}
+ \sum_{\alpha=3}^D \Lambda_{\ldots\alpha\alpha}l_\alpha  
+ \sum_{\alpha,\beta=3}^D 
\left(
1 - \delta_{\alpha\beta} + \frac{1}{2!}\delta_{\alpha\beta}
\right)
\Lambda_{\ldots\alpha\alpha\beta\beta} l_\alpha l_\beta
+ \ldots,                                              
\end{equation}
where taking into account only the first term corresponds to MFA.
These series, describing the influence of pair-correlated fluctuations of 
different components of the molecular field, can be rewritten as
\begin{equation}\label{rencum1}
\tilde\Lambda_{\ldots}
= \prod_{\alpha=3}^D \sum_{n_\alpha=0}^\infty
\frac{1}{n_\alpha !}
\left(
l_{\alpha i}\frac{\partial^2}{\partial \xi_\alpha^2}
\right)^{n_\alpha}
\Lambda_{\ldots}(\mbox{\boldmath$\xi$}_i)
= \exp
\left[
\sum_{\alpha=3}^D 
l_{\alpha i}\frac{\partial^2}{\partial \xi_\alpha^2}
\right]
\Lambda_{\ldots}(\mbox{\boldmath$\xi$}_i) .
\end{equation}
Such exponential differential operators were 
considered by Horwitz and Callen \cite{horcal61} for the Ising model. 
Generalization of their results for the multi-component case yields
a closed formula
\begin{equation}\label{rencum2}
\tilde\Lambda_{\ldots} 
 = \frac{1}{\pi^{(D-2)/2}}\!\!\int\!\!d^{D-2} r \;{\rm e}^{-r^2} 
\Lambda_{\ldots}(\mbox{\boldmath$\zeta$}_i),
\end{equation}
where $\mbox{\boldmath$\zeta$}_i$ is the spread molecular field given by
\begin{equation}\label{sprmf}
\mbox{\boldmath$\zeta$}_i \equiv \mbox{\boldmath$\xi$}_i
+ 2\sum_{\alpha=3}^D l_{\alpha i}^{1/2}r_\alpha {\bf e}_\alpha ,
\end{equation}
and the integration in (\ref{rencum2}) is performed with respect to the 
$(D-2)$-component vector variable ${\bf r} \equiv \{r_\alpha\}$.

The Dyson equation for the spin-spin correlation function 
$S_{jj'}^{\alpha\alpha}$ entering (\ref{lalphai}) is represented in
Fig. \ref{sdwf1}b and has the analytical 
form
\begin{equation}\label{dysoncf}
S_{jj'}^{\alpha\alpha} = 
  \tilde\Lambda_{\alpha\alpha j} \delta_{jj'}
+ \tilde\Lambda_{\alpha\alpha j} \eta_\alpha \beta 
  \sum_r J_{jr}S_{rj'}^{\alpha\alpha} .
\end{equation}
Applying this equation two times, one can simplify the expression 
for the quantity $l_{\alpha i}$ to 
\begin{equation}\label{lalphai2}
l_{\alpha i} = \frac{1}{2\tilde\Lambda_{\alpha\alpha i}}
\left(
\frac{S_{ii}^{\alpha\alpha}}{\tilde\Lambda_{\alpha\alpha i}} - 1
\right) .
\end{equation}
The system of SCGA equations for a domain wall in a biaxial 
ferromagnet stated above simplifies 
for a small anisotropy ($1-\eta \ll 1$) and in the limit $D\to\infty$, 
which will be pursued in the next section. 
For $\eta_\alpha=0$ the 
molecular field fluctuations measured by the quantity $l_{\alpha i}$ 
vanish according to (\ref{dysoncf}) and (\ref{lalphai2}), and the 
magnetization equation (\ref{scgam}) reduces to the MFA result 
(\ref{cweiss}).

\section{The spherical limit}
\eqreset

In the spherical limit $D\to\infty$ the Langevin function $B(\xi)$
(see (\ref{cweiss}), (\ref{cum12}))
simplifies to
\begin{equation}\label{bdgreat}
B(\xi) \cong \frac{x}{1+\sqrt{1+x^2} }, 
\end{equation}
where the scaled variable $x$ 
(which should not be confused with the coordinate $x$ appearing below)
is given by $x\equiv 2\xi/D$. 
Correspondingly, the spin cumulant averages (\ref{defcum}) 
considered as functions of $x$ can be estimated as
\begin{equation}\label{cumddep}
\Lambda_{\alpha_1\alpha_2\ldots\alpha_k} \propto D^{1-k} .
\end{equation}
With the help of this estimate it can be shown that in the limit 
$D\to\infty$ SCGA becomes exact, since all other 
more complicated diagrams die out, at least as $1/D$ \cite {gar94jsp}.
Indeed, a unification of two oval blocks into a larger one, which leads 
to a more complicated diagram (e.g., 
$\Lambda_{\alpha\alpha}\times\Lambda_{\beta\beta}\Rightarrow
\Lambda_{\alpha\alpha\beta\beta}$ in Fig. \ref{sdwf1}), leads to the 
appearance of an additional factor $1/D$, since
\begin{equation}\label{cumcomp}
\Lambda_{\alpha_1\alpha_2\ldots\alpha_{m+n}} \propto 
\Lambda_{\alpha_1\alpha_2\ldots\alpha_m}
\times\Lambda_{\alpha_{m+1}\ldots\alpha_{m+n}}/D .
\end{equation}
In \cite{gar94jsp} some of such higher-order diagrams were 
considered in the framework of the $1/D$ expansion for low-dimensional 
classical ferro- and antiferromagnets.

For the consideration of the limit $D\to\infty$
it is convenient to introduce the well behaved 
dimensionless temperature variable 
$\theta\equiv T/T_c^{\rm MFA}$, where $T_c^{\rm MFA}=J_0/D$ and $J_0$ is the  
zero component of the exchange interaction, as well as the further 
$D$-independent variables:
\begin{equation}\label{dvars}
G_i \equiv (D/\theta)\tilde\Lambda_{\alpha\alpha i}, \qquad
\tilde l_i \equiv l_{\alpha i}/D, \qquad
s_{ii'} \equiv D S_{ii'}^{\alpha\alpha} .
\end{equation}
Expression (\ref{lalphai2}) can now be rewritten as
\begin{equation}\label{lalphai3}
\tilde l_i = \frac{1}{2\theta G_i}
\left( \frac{s_{ii}}{\theta G_i} - 1 \right) ,
\end{equation}
and the expression for the square of the spread value of the
argument $x$ in (\ref{rencum2}) reads
\begin{equation}\label{xspread}
x_i^2 \equiv 
\left( 
\frac{2\mbox{\boldmath$\zeta$}_i}{D} 
\right)^2 
= 
\left(
\frac{2}{\theta} \sum_j \lambda_{ij}m_{zj}
\right)^2
+ 
\left(
\frac{2\eta}{\theta} \sum_j \lambda_{ij}m_{yj}
\right)^2
+ 
\frac{16 \tilde l_i}{D}\sum_{\alpha=3}^D r_\alpha^2 ,
\end{equation}
where $\lambda_{ij}\equiv J_{ij}/J_0$.
It can be seen that the contributions of the fluctuations of the 
$\alpha$-components of the molecular field to 
(\ref{xspread}), each of them being small as 
$1/D$, are essential due to their large number.
Now for $D\gg 1$ the Gaussian integrals (\ref{rencum2}) can be  
easily calculated by applying the identity
\begin{equation}\label{gauident}
\frac{1}{\pi^{1/2}} \!\!\int\limits_{-\infty}^\infty \!\!
dx\;{\rm e}^{-x^2}\! f(ax^2) \cong f(a/2), \qquad  a\ll 1
\end{equation}
successively $D-2$ times. 
Thus, the integration leads simply to the 
replacement $r_\alpha^2 \Rightarrow 1/2$ in (\ref{xspread}). 
Now for the quantity $G_i$ of (\ref{dvars}) with the use of the second of 
the formulae (\ref{cum12}) and of the asymptotic expression (\ref{bdgreat}) 
one gets
\begin{equation}\label{gx}
G_i = \frac{2}{\theta}\frac{1}{1+\sqrt{1+x_i^2} } .
\end{equation}
The magnetization equation (\ref{scgam}) can be simplified by using  
the first of the formulae (\ref{cum12}), as well as (\ref{bdgreat}), to
\begin{equation}\label{sphermageq}
m_{zi} = G_i \sum_j \lambda_{ij}m_{zj}, \qquad
m_{yi} = \eta G_i \sum_j \lambda_{ij}m_{yj} .
\end{equation}
Finally, determining $x_i^2$ from (\ref{gx}) as a function of $G_i$, 
equating it to (\ref{xspread}) with $r_\alpha^2 = 1/2$ and using 
(\ref{lalphai3}) and (\ref{sphermageq}), one arrives at the equation 
\begin{equation}\label{spherelleq}
s_{ii} + m_i^2 = 1 ,
\end{equation}
which is nothing but the kinematic identity ${\bf m}_i^2=1$ 
in the limit $D\to\infty$.
The normalized correlation function $s_{ii'}$ determined by 
(\ref{dvars}) satisfies the linear equation following from (\ref{dysoncf}),
\begin{equation}\label{dysoncf2}
s_{ii'} 
= 
\theta G_i \delta_{ii'}
+ 
\eta_\alpha G_i \sum_j \lambda_{ij}s_{ji'}
\end{equation}
with the variable coefficient $G_i$.

Equations (\ref{sphermageq})--(\ref{dysoncf2}) 
constitute the closed system of equations, which can be 
applied for the calculation of  
the domain wall magnetization profile in the spherical limit.

In the homogeneous case (or in one of the domains) $m_y=0$ and $m_z$ and $G$ 
are constants. 
In this case equation (\ref{dysoncf2}) can be easily 
solved with the help of the Fourier transformation, which results in
\begin{equation}\label{cfhomo}
s_{ii} = 
v_0\!\!\!\int\!\!\!\frac{d{\bf q}}{(2\pi)^3} s_{\bf q} 
=
\theta G P(\eta_\alpha G) , 
\qquad P(X) \equiv 
v_0\!\!\!\int\!\!\!\frac{d{\bf q}}{(2\pi)^3}
\frac{1}{1-X\lambda_{\bf q}} ,
\end{equation}
where
$v_0$ is the unit cell volume and $\lambda_{\bf q} \equiv J_{\bf q}/J_0$. 
In the long-wavelength limit $\lambda_{\bf q} \cong 1 - \alpha q^2$, 
where $\alpha \sim a_0^2$ and $a_0$ is the lattice spacing.
The lattice integral $P(X)$ has the following properties:
\begin{equation}\label{plims}
P(X) \cong \left\{
\begin{array}{ll}
1 + X^2/z,                       & X   \ll 1               \\
W - c_0\,(1-X)^{1/2},            & 1-X \ll 1 ,  
\end{array} 
\right. 
\end{equation}
where $z$ coincides with the number of the nearest neighbors
for the nn interactions 
and $W$ (the Watson integral) and $c_0=v_0/(4\pi\alpha^{3/2})$ 
are lattice-dependent constants. 
For a simple cubic (sc) lattice 
$v_0 = a_0^3$ and $\alpha = a_0^2/6$, hence $c_0 = (2/\pi)(3/2)^{3/2}$.
Since the sum in the right-hand part of the first of the equations 
(\ref{sphermageq}) equals $m_z$, this equation is satisfied only if 
$m_z=0$ (above $T_c$) or $G=1$ (below $T_c$). 
In these cases from  
equation (\ref{spherelleq}) one gets the temperature-dependent 
magnetization $m\equiv m_z$:
\begin{equation}\label{mspher}
m = (1-\theta/\theta_c)^{1/2}, \qquad 
\theta \le \theta_c \equiv 1/P(\eta_\alpha) .
\end{equation}
It can be seen that in the fully isotropic case ($\eta=\eta_\alpha=1$) 
the value of the phase transition temperature in the bulk 
$\theta_c=1/P(\eta_\alpha)$ reduces to the well-known result 
$\theta_c=1/W$ \cite{berkac52}.

The width of a Bloch wall $\delta_B$ in a uniaxial ferromagnet is determined 
by the balance between the anisotropy and inhomogeneous exchange 
energies. 
For small anisotropy ($1-\eta\ll 1$) the condition 
$\delta_B \gg a_0$ is fulfilled.
In this case the change of the domain 
wall magnetization at a distance of $a_0$ is small, 
and for systems with nn interactions one can rewrite 
the lattice sum in (\ref{sphermageq}) about some point 
${\bf r}\equiv {\bf r}_i$ as
\begin{equation}\label{laplmag}
\sum_j \lambda_{ij}{\bf m}_j \cong {\bf m}({\bf r}) 
+ 
\alpha\Delta{\bf m}({\bf r}) ,
\end{equation}
where the second term with the Laplace-operator $\Delta$ is small in 
comparison to the first one.
The quantity $G_i$ in (\ref{sphermageq}) 
determined by (\ref{dvars}) is in the small-anisotropy case 
also a continuous 
function of the coordinate with the scale $\delta$. 
Moreover, as can be 
seen from (\ref{sphermageq}) and (\ref{laplmag}), the deviation of $G$ 
from its bulk-value 1 is small, i.e.,
\begin{equation}\label{g1def}
G({\bf r}) = 1 + G_1({\bf r}),  \qquad  G_1 \ll 1 .
\end{equation}
Now one can 
rewrite the equations (\ref{sphermageq}) in terms of the normalized 
magnetization ${\bf n}\equiv {\bf m}/m_e$, where $m_e$ is the equilibrium 
bulk magnetization given by (\ref{mspher}). 
The result in the chosen 
geometry is of the form
\begin{eqnarray}\label{sphermageq2}
&&
\alpha n_z''(x) = -G_1(x)n_z(x) , \\
&&
\alpha n_y''(x) = -G_1(x)n_y(x) + (1-\eta)n_y(x) .
\nonumber
\end{eqnarray}
The kinematic equation (\ref{spherelleq}) can be represented in terms of 
${\bf n}$ as 
\begin{equation}\label{spherelleq2}
\epsilon (1-n^2) = \theta_c s_{ii}/\theta - 1, \qquad
\epsilon \equiv \theta_c /\theta - 1 .
\end{equation}
Unlike the magnetization equation (\ref{sphermageq}), the equation for 
the correlation function (\ref{dysoncf2}) cannot, in general, be written in 
a continuous form of the type (\ref{sphermageq2}). 
In the general case 
we are going to consider, where $\eta_\alpha$ is not necessarily 
close to 1, the correlation length of the $\alpha$ spin components
below $\theta_c$,
\begin{equation}\label{rcalpha}
\xi_{c\alpha}=\sqrt{\alpha\eta_\alpha/(1-\eta_\alpha)} 
\end{equation}
(not to be confused with components of the normalized molecular field 
$\mbox{\boldmath$\xi$}$),
which can be determined from $s_{\bf q}$ in (\ref{cfhomo}),
can be comparable with the lattice spacing $a_0$. 
Moreover, even in the case $1-\eta_\alpha \ll 1$ the continuous 
approximation for $s_{ii'}$ does not yield the correct bulk result 
(\ref{cfhomo}) which is formed by the integration 
over the whole 
Brillouin zone and not only over the logwavelength region $q \ll 1$.
But it can, nevertheless, be applied for the calculation of the wall 
properties, as we shall see below.

\section{Domain wall magnetization profile}
\eqreset

Before proceeding to the solution of equations (\ref{sphermageq2}), 
(\ref{spherelleq2}) and (\ref{dysoncf2}) in the general situation, let us 
consider at first the case $\eta_\alpha=0$, where the spin fluctuations 
play no role and the situation is described exactly by MFA. 
Here the 
solution of (\ref{dysoncf2}) yields $s_{ii}=\theta G_i$, and with 
the help of (\ref{spherelleq2}) one gets 
$G_1(x) = \epsilon [1-n^2(x)]$. 
Adopting it in 
(\ref{sphermageq2}) and solving the resulting equations, one arrives at 
the magnetization profile \cite {bulgin64,sartrubis76}
\begin{equation}\label{magprofew}
n_z = \pm \tanh{(x/\delta)}, \qquad  n_y = \pm \rho/\cosh{(x/\delta)} ,
\end{equation}
where
\begin{equation}\label{rhodelta}
\rho =
\left\{
\begin{array}{ll}
\sqrt{1-\tau} ,  & \tau \equiv 2(1-\eta)/\epsilon \le 1 
\\
0 ,              & \tau \ge 1 ,   
\end{array} 
\right. \qquad 
\delta =
\left\{
\begin{array}{ll}
\delta_B = \sqrt{\alpha/(1-\eta)} ,                         & \tau \le 1
\\
\delta_L = \sqrt{2\alpha/\epsilon} = \delta_B \tau^{1/2},   & \tau \ge 1 .
\end{array} 
\right. 
\end{equation}
It can be seen that the crossover from the Bloch ($\rho=1$) to the linear 
($\rho=0$) wall proceeds with increasing temperature 
through the elliptic one having 
$n_z^2 + n_y^2/\rho^2 = 1$ with $0<\rho<1$, and the transverse component of 
the magnetization in the wall, $n_y$, disappears through a second-order
phase transition.
The temperature of the DW phase transition can be written in three forms:
\begin{equation}\label{critparmfa}
\tau_B = 1, \qquad  \epsilon_B = 2(1-\eta) \ll 1,  \qquad
\theta_B = 1/(1+\epsilon_B) \cong 1 - \epsilon_B ,
\end{equation}
the quantity $\tau$ playing the same role for a DW as 
the ``absolute'' temperature $\theta$ for the bulk (cf. (\ref{mspher})).
The temperature-dependent factor $\rho$ in 
(\ref{magprofew}) can be interpreted as the DW order parameter $m_B$
\cite {koegarharjah93,harkoegar95}. 
Whereas the Bloch-wall width
$\delta_B$ is temperature-independent, the width of the linear wall 
$\delta_L$ is determined by 
the balance of the homogeneous and inhomogeneous exchange energies and is 
diverging at $\theta_c$. 
Considering the first of equations 
(\ref{magprofew}) for $x \gg \delta_L$, one can identify
\begin{equation}\label{identxi}
1-\tanh(x/\delta_L) \cong \exp(-x/\xi_{cz}) \Longrightarrow
\delta_L = 2\xi_{cz} ,
\end{equation}
where $\xi_{cz}$ is the correlation length of the $z$ spin components.
One should also note the analogy between the Bloch-wall width $\delta_B$
(\ref{rhodelta}) and the transverse correlation length $\xi_{c\alpha}$ 
(\ref{rcalpha}), which coinside for a purely uniaxial 
($\eta_\alpha=\eta$) model with small anisotropy.
The function $G_1$ entering equations (\ref{sphermageq2}) can be 
written as
\begin{equation}\label{g1}
G_1(x)  
= 
\frac{\epsilon (1-\rho^2)}{\cosh^2{(x/\delta)}}
=
\frac{2\alpha}{\delta^2}\frac{1}{\cosh^2{(x/\delta)}} .
\end{equation}
Since $\alpha\sim a_0^2$, in the small-anisotropy case
$G_1 \sim (a_0/\delta)^2 \ll 1$ in the whole temperature interval.

Now we proceed to the solution of the magnetic interface problem 
described by equations (\ref{sphermageq2}), (\ref{spherelleq2}), and 
(\ref{dysoncf2}) in the general case $\eta_\alpha \ne 0$. 
The solution of the Dyson 
equation for the correlation function (\ref{dysoncf2}) depends on the 
relation between the correlation length $\xi_{c\alpha}$ of (\ref{rcalpha}) 
and the other length scales, $a_0$ and $\delta$. 
If $\xi_{c\alpha} \gg a_0$, which is satisfied for $1-\eta_\alpha \ll 1$, 
the continuous approximation to equation (\ref{dysoncf2}) can be 
applied. 
If $\xi_{c\alpha} \ll \delta$ 
(i.e., $1-\eta_\alpha \gg 1-\eta$, see (\ref{rhodelta})), the correlation 
function $s_{ij}$ can be easily calculated {\em locally} with respect 
to the slowly changing magnetization profile 
(or, more exactly, the profile of $G$) in the wall. 
For uniaxial ferromagnets with a small anisotropy 
($1-\eta \ll 1$) considered throughout this paper these limiting cases 
overlap in the region $1-\eta \ll 1-\eta_\alpha \ll 1$. 
Let us consider at first the case $\xi_{c\alpha} \ll \delta$. 
Here one can make a replacement 
$G_i \Rightarrow G_{i'}$ in (\ref{dysoncf2}), after which this 
equation can be solved like in the homogeneous case. 
With the use of (\ref{cfhomo}) 
and (\ref{g1def}) one gets
\begin{equation}\label{cfquasihomo}
s_{ii} 
\cong 
\theta G_i P(\eta_\alpha G_i)
\cong
\theta P(\eta_\alpha) [1 + I(\eta_\alpha) G_1(x)] ,
\end{equation}
where
\begin{equation}\label{ietaalpha}
I(\eta_\alpha) = 1 + \frac{\eta_\alpha P'(\eta_\alpha)}{P(\eta_\alpha)}
\cong
\left\{
\begin{array}{ll}
\displaystyle
1 + 2\eta_\alpha^2/z ,       & \eta_\alpha \ll 1
\\
\displaystyle
\frac{c_0}{2P(\eta_\alpha)}\frac{1}{\sqrt{1-\eta_\alpha}} , 
& 1-\eta_\alpha \ll 1 
\end{array}
\right.
\end{equation}
and $c_0$ is determined by (\ref{plims}).
Now with the use of (\ref{cfquasihomo}) and (\ref{spherelleq2}) one can 
express $G_1$ through the magnetization profile $n(x)$:
\begin{equation}\label{g1ren}
G_1(x) = \epsilon [1-n^2(x)] I^{-1}(\eta_\alpha) .
\end{equation}
This expression differs only by a constant from that used in the 
beginning of this section in the MFA limit $\eta_\alpha=0$. 
Solving now the magnetization equations 
(\ref{sphermageq2}) like above, one gets the same  DW magnetization 
profile (\ref{magprofew}), where the parameters $\rho$ and $\delta$ 
are given by (\ref{rhodelta}) with the renormalized DW temperature: 
$\tau\Rightarrow\tilde\tau \equiv \tau I(\eta_\alpha)$.
The critical values of the 
three temperature variables (cf. (\ref{critparmfa})) read now 
\begin{equation}\label{critpar}
\tau_B = I^{-1}(\eta_\alpha) < 1, \qquad  
\epsilon_B = 2(1-\eta)I(\eta_\alpha) ,  
\qquad
\theta_B = \theta_c/(1+\epsilon_B) ,
\end{equation}
where $\theta_c$ is given by (\ref{mspher}). 
One can see from 
(\ref{critpar}) and (\ref{ietaalpha}) that for 
$1-\eta_\alpha \ll 1$ the effective transition temperature $\tau_B$ 
becomes small. 
On the other hand, due to the validity condition 
$1-\eta_\alpha \gg 1-\eta$ the absolute temperature $\theta_B$ of the DW 
phase transition remains in the limiting case under consideration close to 
$\theta_c$ ($\epsilon_B \ll 1$). 
It can be seen that in this case the domain wall does not 
demonstrate any 2-dimensional behaviour, and its phase transition 
at $\theta_B < \theta_c$ can still be described qualitatively 
by the effective diminishing of the 
ordering interaction for the wall spins forced perpendicularly to the 
easy axis $z$, as was said in the Introduction. 
The effect in the case $1-\eta_\alpha \ll 1$ is much larger than according to
the MFA estimates because of the 
divergence of the function $I(\eta_\alpha)$ of (\ref{ietaalpha}).
Accordingly, the width of the linear wall can considerably exceed its
mean-field value (\ref{rhodelta}):
\begin{equation}\label{deltal}
\delta_L = \sqrt{ \frac{2\alpha}{\epsilon} }\cdot I^{1/2}(\eta_\alpha)
\cong
\sqrt{ \frac{\alpha c_0}{\epsilon P(\eta_\alpha)} } 
\frac{1}{(1-\eta_\alpha)^{1/4}} 
\end{equation}
in the case $1-\eta_\alpha \ll 1$.
Since $\delta_L$ is related to the longitudinal correlation length
(see (\ref{identxi})), this result shows a non-trivial influence of 
fluctuations of transverse spin components on the longitudinal spin 
correlations in the anisotropic spherical model.
One can also check that the function 
$G_1(x)$ of (\ref{g1ren}) is still given by the expression (\ref{g1}) 
with the changed value of the DW width $\delta$.

In the other limiting case, $\xi_{c\alpha} \gg a_0$, a continuous 
approximation of the type (\ref{laplmag}) can be applied to the Dyson 
equation (\ref{dysoncf2}). 
Making the Fourier transformation with 
respect to the coordinates $y$ and $z$ and using the conditions 
$1-\eta_\alpha \ll 1$ and (\ref{g1def}) one arrives at the
differential equation for the correlation function
\begin{equation}\label{diffcf}
\alpha s''(x) - [1-\eta_\alpha + \alpha q^2 - G_1(x)]s(x)
= -\theta \delta(x-x') ,
\end{equation}
where $q^2\equiv q_y^2+q_z^2$ and the ``mute'' argument $x'$ of $s$ 
was dropped. 
This equation should be solved to yield $s$ with 
$x=x'$ as a function or functional of $G_1$, and $s_{ii}$ 
(cf. (\ref{cfquasihomo})) should be obtained by the integration 
of $s$ over $q_y$ and $q_z$. 
Then, as above, $G_1$ should be found 
from (\ref{spherelleq2}) and used in the magnetization equations 
(\ref{sphermageq2}). 
All this seems to be too complicated since 
equation (\ref{diffcf}) cannot be solved analytically for the arbitrary 
function $G_1(x)$. 
But the expected result that the DW transition 
temperature $\theta_B$ turns to zero in the purely uniaxial case 
$\eta_\alpha = \eta$ signals that there should be an exact solution 
to the problem. 
We can try to find it assuming that
$G_1(x)$ has the same functional form as above, (\ref{g1}), 
with some renormalized value of the DW width $\delta$ as a parameter. 
Then using a new variable $u\equiv \tanh(x/\delta)$, 
equation (\ref{diffcf}) can be rewritten as
\begin{equation}\label{diffcfu}
\frac{d}{du}(1-u^2)\frac{ds}{du} + 
\left( 2 - \frac{\mu^2}{1-u^2} \right)\! s(u) = 
-\frac{\delta\theta}{\alpha}\delta(u-u') , \qquad
\mu^2 \equiv \frac{\delta^2}{\alpha}(1-\eta_\alpha + \alpha q^2) 
\end{equation}
and solved in terms of the adjoined Legendre functions
\begin{equation}\label{legendre}
{\rm P}_1^{\pm\mu}(u) = \frac{u\pm\mu}{\Gamma(2\pm\mu)}
\left( \frac{1+u}{1-u} \right)^{\pm\mu/2} ,
\end{equation}
which leads to
\begin{equation}\label{cfgen}
s(x,x',q) 
= 
\frac{\theta\delta}{2\alpha\mu} 
\exp\left( -\frac{\mu}{\delta}|x-x'| \right)
\left[
1 + \frac{1-\tanh\frac{x}{\delta}\tanh\frac{x'}{\delta}}{\mu^2-1}
\left( 1 + \mu\tanh\frac{|x-x'|}{\delta} \right)
\right] .
\end{equation}
This (not translationally invariant) expression can be reduced 
in the case $x=x'$ with the help of (\ref{g1}) to the form
\begin{equation}\label{cfx}
s(x,x,q) = 
\frac{\theta}{2\alpha^{1/2}\sqrt{1-\eta_\alpha + \alpha q^2}}
\left[ 
1 + \frac{G_1(x)}{2[1-\eta_\alpha - (1-\eta)/\tilde\delta^2 + \alpha q^2]}
\right] ,
\end{equation}
where $\tilde\delta \equiv \delta/\delta_B$ (see (\ref{rhodelta})). 
Since the structure of this expression is analogous to that of 
(\ref{cfquasihomo}), it is now clear that the choice of $G_1$ in the 
form (\ref{g1}) was right. 
Integrating (\ref{cfx}) over the 2-dimensional 
wavevector ${\bf q}$ to get $s_{ii}$ and proceeding as above, one gets 
functionally the same results (\ref{magprofew}) and (\ref{rhodelta}) with 
a new renormalized DW temperature
\begin{equation}\label{tautilde}
\tilde\tau \equiv \tau I(\eta_\alpha,\eta,\tilde\delta) ,
\end{equation}
where 
\begin{equation}\label{ietaalpha2}
I(\eta_\alpha,\eta,\tilde\delta) 
= 
\frac{c_0}{4P(\eta_\alpha)} \frac{\tilde\delta}{\sqrt{1-\eta}}
\ln \frac
{ \tilde\delta\sqrt{1-\eta_\alpha} + \sqrt{1-\eta} }
{ \tilde\delta\sqrt{1-\eta_\alpha} - \sqrt{1-\eta} }  .
\end{equation}
The latter simplifies in the limit $1-\eta_\alpha \gg 1-\eta$ 
to the second limiting expression in (\ref{ietaalpha}).
It can be seen from (\ref{cfx}) that 
in the case under consideration, $1-\eta_\alpha \ll 1$,
the integral (\ref{ietaalpha2})
is determined by the long-wavelength region, which justifies using the 
continuous approximation for the transverse correlation function $s$.
In the region of elliptic walls ($\tilde\tau < 1$) one has 
$\delta=\delta_B$ 
and hence in (\ref{ietaalpha2}) one has $\tilde\delta=1$. 
The critical values of 
both the DW temperature $\tau$ and the absolute temperature $\theta$ 
(\ref{critpar}) determined now by $I(\eta_\alpha,\eta,1)$ 
go to zero in the uniaxial limit:
\begin{equation}\label{critpar2}
\tau_B \propto \theta_B \propto 1\bigg/\ln\frac{1}{\eta-\eta_\alpha},
\qquad \eta_\alpha \to \eta ,
\end{equation}
which corresponds to the 2-dimensional nature of domain walls. 
One can see that the 2-dimensional effects lead to a further 
decrease of the 
DW phase transition temperature: $\tau_B\equiv I^{-1}(\eta_\alpha,\eta,1)
< \tau_B^{(0)}\equiv I^{-1}(\eta_\alpha)$, 
where $I(\eta_\alpha)$ is given by (\ref{ietaalpha}).

In the linear-wall region ($\tilde\tau > 1$) the normalized wall width 
$\tilde\delta$ is given 
by the solution of the transcendental equation 
$\tilde\delta^2 = \tau I(\eta_\alpha,\eta,\tilde\delta)$  
following from (\ref{rhodelta}) and (\ref{tautilde}). 
The latter can be rewritten in the natural units in the form
\begin{equation}\label{deltaeq2}
\delta_L = \frac{ \sqrt{\alpha} }{2\epsilon}\frac{c_0}{P(\eta_\alpha)}
\ln\frac{\delta_L + \xi_{c\alpha} }{\delta_L - \xi_{c\alpha} }, 
\end{equation}
where $\xi_{c\alpha}$ is the transverse correlation length given by
(\ref{rcalpha}). 
Far from $\tau_B$, where $\delta_L \gg \xi_{c\alpha}$, 
the solution of (\ref{deltaeq2}) leads to the formula (\ref{deltal}). 
This asymptotic dependence can also be represented in the form 
$\tilde\delta=\sqrt{\tau I(\eta_\alpha)}=\sqrt{\tau/\tau_B^{(0)}}$,
which is the analogue of Curie-Weiss asymptote for the bulk 
susceptibility $\chi(T)$ of a ferromagnet far above $T_c$.
In the purely isotropic model, $\eta_\alpha=\eta$, the linear 
DW structure is realized in the whole temperature range. 
The DW width 
$\delta_L$ determined by the solution of (\ref{deltaeq2}) shows a 
crossover to the Bloch-wall width $\delta_B$ at low temperatures:
\begin{equation}\label{deltalb}
\delta_L \cong \delta_B 
\left[
1 + 2\exp
\left(
-\frac{2\epsilon}{\sqrt{1-\eta} }\frac{P(\eta)}{c_0}
\right)
\right], \qquad \epsilon \equiv \theta_c/\theta -1 \gg \sqrt{1-\eta} .
\end{equation}
It is worth to note that in this limiting case the longitudinal 
correlation length, $\xi_{cz}=\delta_L/2$, is determined by the 
transverse one, $\xi_{c\alpha}=\delta_B$.

\begin{figure}[t]
\unitlength1cm
\begin{picture}(15,8)
\centerline{\epsfig{file=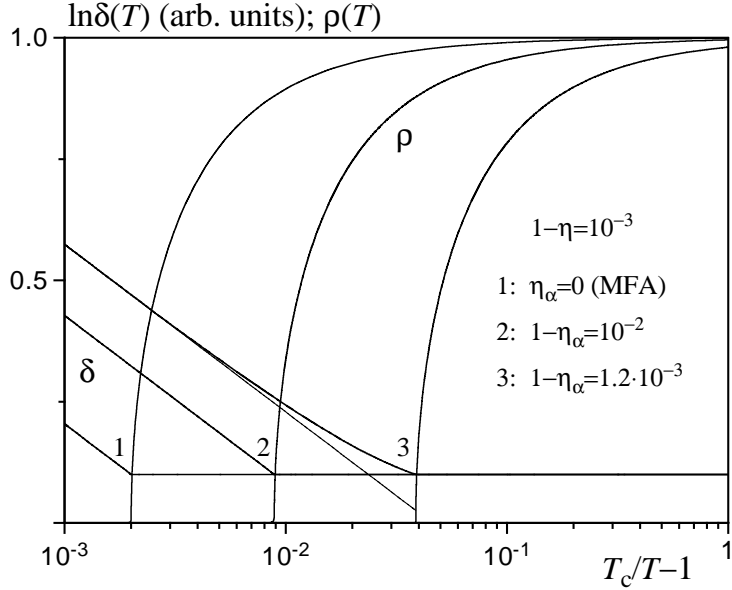,angle=-90,width=15cm}}
\end{picture}
%
\caption{ \label{rhodeltavseps}
Temperature dependences of the domain wall order parameter 
$m_B\equiv\rho$ and the DW width $\delta$ 
for different values of the in-plane anisotropy parameter $\eta_\alpha$.
}
\end{figure}

The temperature dependences of the DW order parameter $m_B\equiv \rho$
and the normalized DW width $\tilde\delta$ are represented in 
Fig. \ref{rhodeltavseps} for different values of $\eta_\alpha$ and 
$1-\eta=10^{-3}$. 
One can see that for $1-\eta_\alpha = 10^{-2}$ the 
fluctuational decrease of the DW transition temperature is essential, 
although 2-dimensional effects are still negligible in this case. 
In contrast, for $1-\eta_\alpha = 1.2\cdot 10^{-3}$ they come into play, 
the corresponding additional diminishing of the transition
temperature becomes essential, and 
the temperature dependence of $\tilde\delta$ in the logarithmic scale 
is no longer a straight line. 

\par

\section{Discussion}
\eqreset

In the main part of this paper the structure of domain walls in the 
biaxial ferromagnetic model described by the classical-vector Hamiltonian 
(\ref{dham}) was calculated exactly in the whole temperature range 
$T<T_c$ in the spherical limit $D\to\infty$. 
The main qualitative result 
is that in the purely uniaxial model ($\eta_\alpha=\eta$) the Bloch walls 
do not exist at any non-zero temperature (see (\ref{critpar2})) 
being disordered by thermal fluctuations to the linear (Ising-like) ones.  
This result complements the well-known fact that the Bloch walls 
in the purely uniaxial model (considered within the phenomenological 
micromagnetic approach which is equivalent to MFA) cannot move 
since their maximal velocity (the Walker velocity) is equal to zero. 
In the opposite limit, $\eta_\alpha=0$, the model with $D\to\infty$ 
{\em total} spin components and a finite number (here 2) of 
{\em interacting} ones is realized. 
In this case spin fluctuations die out and MFA becomes exact. 
The temperature of the phase transition 
from Bloch to linear walls $T_B$ changes as function of $\eta_\alpha$ 
from its MFA value $T_B=(1-\epsilon_B)T_c$, 
$\epsilon_B=2(1-\eta) \ll 1$ at 
$\eta_\alpha=0$ to 0 at $\eta_\alpha=\eta$ 
(see (\ref{critparmfa}) and (\ref{critpar})).

In the actual case of a small anisotropy, $1-\eta \ll 1$, the behavior 
of a domain wall is more complicated than that of a purely 2-dimensional 
object, since the DW width $\delta$ is much larger than the lattice 
spacing $a_0$. 
In the spherical limit this leads to the existence of two 
different mechanisms of the DW ordering, depending on the value of 
$\eta_\alpha$. 
For the strong anisotropy in the basis plane $x,y$, i.e., 
$1-\eta_\alpha \gg 1-\eta$, the correlation length $\xi_{c\alpha}$ of 
(\ref{rcalpha}) for all temperatures is much shorter than the wall width 
$\delta$ in (\ref{rhodelta}), and the DW phase transition at $T_B$ can be 
interpreted as the locally shifted bulk one (the ``perturbed 
3-dimensional'' case, see (\ref{cfquasihomo})--(\ref{critpar})). 
In contrast, in the case of the two anisotropies comparable with each other, 
$1-\eta_\alpha \sim 1-\eta$, the true 2-dimensional situation is realized
(see (\ref{ietaalpha2}) and (\ref{critpar2})). 
Such a separation does not, 
however, persist for models with finite values of $D$ (e.g., for the 
Heisenberg model, $D=3$), where the DW phase transition should always be
of a 2-dimensional character. 
For such models the fluctuations of the ordering 
spin components, $m_z$ and $m_y$, also play a role, and in the 
temperature interval about $T_B$, where the diverging correlation 
length $\xi_{cy}$ exceeds the DW width $\delta$, a 2-dimensional 
behavior is realized. 
Since $\delta \gg a_0$, this temperature interval should be much 
narrower than for a purely 2-dimensional system. 
The asymptotic critical 
behavior of the DW order parameter $m_B=\rho$ 
in (\ref{magprofew}) is described by the 
critical index $\beta_B=1/8$ of the 2-dimensional Ising model, as was 
confirmed experimentally in \cite {koegarharjah93,harkoegar95}.

As a subject of future investigations, the temperature dependence of the 
DW magnetization profile in the self-consistent Gaussian approximation, 
without going to the limit $D\to\infty$, can be considered. 
Although it 
can be connected with more complicated numerical calculations, one can 
expect to obtain, with the help of SCGA, essentially more accurate results 
for domain walls in comparison with the spherical approximation, as was 
demonstrated for the bulk properties \cite {garlut86jpf,gar96prb}. 
The other problem is to formulate dynamic equations for fluctuating domain 
walls and to calculate their mobility. 
Such equations can, in principle, 
be obtained with the help of some dynamical generalization of the 
classical spin diagram technique \cite {garlut84d}. 
Unfortunately, in the 
dynamical case one cannot make use of going to the limit $D\to\infty$ 
with all related simplifications, and only SCGA for the Heisenberg model 
can be used as the underlying static approach.

A promising field for the application of the methods of this paper is the 
surface effects in finite and semi-infinite magnetic systems,
which are very sensitive to the anisotropy. 
This problem was addressed recently in \cite{gar96jpal}.

\section*{Acknowledgements}

The author thanks Hartwig Schmidt and J.K\"otzler for valuable discussions
and Scott Allen, whose careful study of the manuscript 
helped to eliminate some inaccuracies. 

The financial support of Deutsche Forschungsgemeinschaft 
under contract Schm 398/5-1 is greatfully acknowledged.


\end{document}